\documentstyle[prl,aps]{revtex}

\begin{document}

\draft
\title{\vskip -5cm \begin{flushright} {\bf BROWN HET-1070} 
\end{flushright} \vskip 2cm  
Origin of Magnetic Fields in the Universe Due to Nonminimal 
Gravitational-Electromagnetic Coupling \thanks{ To appear in Physical 
Review Letters}}
\author{Reuven Opher$^{1}$ \cite{A1}, 
Ubirajara F. Wichoski$^{2, 3}$ \cite{A2} \vspace*{2mm} }

\address{$^1$Instituto de Astronomia e Geof\'isica, Universidade 
de S\~ao Paulo, \\ Av. Miguel Stefano, 4.200, CEP 04301-904, 
S\~ao Paulo, SP, Brazil
\vspace*{2mm}\\
$^2$Department of Physics, Brown University, Providence, RI 
02912, USA
\vspace*{2mm}\\
$^3$Depto. F\'isica-Matem\'atica, Instituto de F\'isica, 
Universidade de S\~ao Paulo, \\ Caixa Postal 66318,  CEP 05389-970, 
S\~ao Paulo, SP, Brazil}

\date{\today}
\maketitle

\begin{abstract}
Basically the only existing theories for the creation of 
a magnetic field (${\bf B}$) in the Universe are the creation of a seed
field  $\sim 10^{-20}\;\text{G}$ in spiral galaxy which is subsequently
supposedly amplified up to the observed $10^{-6} \text{---} 
10^{-5}\;\text{G}$ by a dynamo process or a seed intergalactic field of
$\sim 10^{-12} \text{---}10^{-10}\;\text{G}$ which is amplified by 
collapse and differential rotation. No satisfactory dynamo theory,
however, exists today. We show that a $\sim 10^{-6} \text{---} 
10^{-5}\;\text{G}$ magnetic field in spiral galaxies is directly
obtained from a nonminimal gravitational-electromagnetic coupling,
without the need of significant dynamo amplification. 
\end{abstract}

As far as we know, cosmic magnetic fields pervade the Universe.
Nevertheless, we still do not know their origin, which 
has been pursued according 
to two basic possibilities, namely cosmological (i.e., 
primordial origin) \cite{kro} and Biermann-type battery seed effects
\cite{bie}.

Several alternatives have been suggested to account for the cosmological 
origin of the seed magnetic field, among them are the following: 
{\it i)} It could arise during 
cosmological phase transitions which took place in the early Universe 
\cite{vac,kib}; {\it ii)} It could be generated during an early epoch 
of inflation in inflationary Universe models \cite{rat}; {\it iii)} 
It could emerge in string cosmology from the amplification of the  
electromagnetic vacuum fluctuations due to a dynamical dilaton 
background \cite{gas}. In any case, however, its strength 
is constrained by the abundances of the 
elements formed in big-bang nucleosynthesis \cite{che}. Later, 
this seed field could possibly be amplified either through a 
dynamo mechanism from an initial strength of 
$\sim 10^{-20}\;\text{G}$ \cite{kro}; or through 
protogalactic collapse and 
differential rotation from a initial strength of
$\sim 10^{-12} \text{---} 10^{-10}\;\text{G}$ \cite{pid,kul}, 
to the presently observed 
$10^{-6} \text{---} 10^{-5}\;\text{G}$ \cite{kro}.

The Biermann-type battery seed effect \cite{bie} is 
based on the fact that a magnetic field could be generated as 
long as electronic temperature and density gradients are not parallel, 
generally in a rotating medium. 
Harrison \cite{har} suggested a pregalactic origin where a cosmic 
battery could operate before the recombination epoch creating a 
weak field. This idea has not been pursued because the primordial 
vorticity required by this mechanism could not be sustained in view 
of whirls decaying during cosmic expansion (see also 
\cite{laz,sil,cha} for a galactic origin of the seed magnetic field in 
early stages of the galaxy formation).  

In the case of a dynamo process, it is generally assumed that the 
seed field $\sim 10^{-20}\;\text{G}$
is subsequently amplified up to the observed $10^{-6} \text{---} 
10^{-5}\;\text{G}$ in a spiral galaxy, i.e., an
amplification of over 14 orders of magnitude. No satisfactory dynamo
theory, however, exists today. 
Field \cite{fie} recently discussed the origin of magnetic 
fields in spiral galaxies starting from a seed field 
which is subsequently amplified by dynamo action. 
The first term of the dynamo equation for the increase of the magnetic
field describes the transformation of a poloidal magnetic 
field into a toroidal magnetic field; the second term, the $\alpha$
term, describes the transformation of the toroidal field back into a
poloidal field; and the third term, the $\beta$ term, describes
turbulent diffusion. 
Field \cite{fie} indicates that the traditional values of $\alpha$ 
and $\beta$ are $\alpha \sim 10^{4}\;\text{cm}\;\text{s}^{-1}$ and 
$\beta \sim 10^{26}\;\text{cm}^{2}\;\text{s}^{-1}$. However, for 
$\alpha = \alpha_{0} \sin{(\pi x h^{-1})}$ [with $h$ the height of the 
disc and $x$ the distance perpendicular to the disc] and 
$\alpha_{0} = 2.1 \times 10^{4}$ and $\beta \cong 10^{26}$, 
we have the growth rate less than zero (i.e., 
no dynamo amplification)! The dynamo action is thus extremely 
sensitive to the exact value of $\alpha$ and $\beta$ used. 
In general, turbulent velocities are assumed to determine $\alpha$ 
and $\beta$, rather than determining $\alpha$ and $\beta$ 
self-consistently (i.e., making sure that the derived $\alpha$ and
$\beta$ create the turbulent velocities assumed). Another problem
\cite{oph} is that it is 
assumed in dynamo theory that we have 
$\sim 10^{10}$ years to amplify the seed field but in high 
redshift ($z$) 
systems there is evidence of ${B} \sim 10^{-6}\;\text{G}$, 
requiring dynamo action possibly in $\sim 10^{9}$ years.

The other mechanism, distinct from the dynamo and probably the 
most likely explanation for the galactic magnetic field to date, 
is the amplification  
of a seed field by anisotropic protogalactic collapse
and differential rotation \cite{pid,kul,pi2,ku2} (and references 
therein). In this case, a seed field $\sim 10^{-12} \text{---} 
10^{-10}\;\text{G}$, frozen into the galactic gas, is needed. 
In particular, for spiral galaxies the field has to be also 
oblique to the rotation vector. Inflation is 
a good candidate to produce a seed field of this magnitude 
(e.g. \cite{rat} and references therein).

We argue in favor of a protogalactic origin for the 
$\sim 10^{-6}\;\text{G}$ 
magnetic field which originates from the angular momentum of the 
protogalaxies through the nonminimal gravitational-electromagnetic 
coupling (NMC) between gravitational and electromagnetic fields. 
Gravitational nonminimal coupling has long been considered in the 
literature \cite{ber,goe,nov}. 
In particular, a lot of work has been done on the nonminimal 
coupling between gravitational and electromagnetic fields. It has 
been motivated in part, by the Schuster-Blackett (S-B) conjecture. 
This conjecture, as A. Schuster \cite{sch} first stated at the 
turn of the century, says that the magnetic fields of 
planets and stars arise only from their rotation. In other 
words, neutral mass currents generate magnetic fields 
implying the existence of a NMC 
between gravitational and electromagnetic fields. 

An early attempt to encompass the S-B conjecture in a gravitational 
theory was made by Pauli \cite{pau} in the 1930's. During the 1940's 
and 1950's, after Blackett \cite{bla} resuscitated the conjecture, 
many authors such as Bennett {\it et al.} \cite{ben}, Papapetrou 
\cite{pap}, and Luchak \cite{luc} also attempted to encompass the S-B 
conjecture in a gravitational theory. Later in the eighties, 
Barut and Gornitz \cite{bar} tried to accomplish this 
objective as well. The majority of these works were based on the 5D 
Kaluza-Klein formalism. This formalism was  
used in order to describe a unified 
theory of gravitation and electromagnetism with NMC in such a way that 
the S-B conjecture would be obtained. 
More recently, De Sabbata and Gasperini \cite{des} proposed a theory 
where the relation between neutral mass currents and magnetic 
fields are due to the initial conditions of the Universe, provided 
that torsion is introduced according to the Einstein-Cartan theory, 
and the large-number hypothesis of Dirac is assumed. 
Wesson \cite{wes} and De Sabbata and Gasperini 
\cite{de2}, based on the relation between magnetic fields and 
angular momentum as implicated by the S-B conjecture, argued that 
there is a possible connection between atomic physics and gravitational 
physics. 

Nonminimal gravitational-electromagnetic coupling 
indicates the relation between the angular momentum ${\bf L}$ 
and the magnetic dipole moment ${\bf m}$ 
\begin{equation}
{\bf m} = \left[ \beta\frac{\sqrt{G}}{2 c}\right] {\bf L} \, , 
\label{e1}
\end{equation}
where $\beta$ is a constant, 
$G$ is the Newtonian constant of gravitation and $c$ is the 
speed of light. It is important to mention that the 
relation (1) is speculative and the observational and experimental 
evidence that
exists on its behalf is still not conclusive. 

The observational and experimental 
effort supporting the S-B conjecture includes 
the early work of Blackett \cite{bla}, Wilson \cite{wil}, and Swann 
and Longacre \cite{swa}. More recently, the observational evidence for 
the S-B conjecture is based on the works of Sirag and Woodward. 
Sirag \cite{sir} compared 
the predictions of Eq.(\ref{e1}) to the observed values of the ratio 
of magnetic moment to angular momentum for the Earth, 
Sun, the star 78 Vir, the Moon, Mercury, Venus, Jupiter, Saturn, and 
the neutron star Her X-1. The minimum data for $\beta$ for these 
objects was: 0.12, 0.02, 0.02, 0.11, 0.37, 0.04, 0.03, 0.03, and 0.07, 
respectively. Excluding the star 78 Vir, the maximum data for $\beta$ 
was 0.77 for the planet Mercury (see also \cite{ahl,wki}). 
Woodward \cite{woo} examined the S-B 
conjecture in the context of pulsar gyromagnetic ratios, for 
short-period pulsars. He found that: 1) $\beta$ is not 
the same for all pulsars; 2) Young pulsars evolve with their 
individual value of $\beta$, constant for a discernible period of 
time; and 3) $\beta$ lies in the range 0.001 to 0.01. In the 
present paper we suggest that $\beta$ for galaxies is in the range 
of 0.01 to 0.1, consistent with the data of Sirag \cite{sir} and 
Woodward \cite{woo}.

We apply the relation (\ref{e1}) to protogalaxies just after they 
acquired angular momentum. 
In order to use this relation for a protogalaxy, we have to 
discuss the origin of its angular momentum.

At the present moment, it is believed that the angular momentum of
galaxies was acquired during the protogalaxy stage through the tidal
torques by neighboring protogalaxies \cite{hoy,pee,whi}. 
The best results are accomplished through N-body simulations 
\cite{whi,efs,pad,ban}. 
In the simulation the angular momentum is written in terms of the 
spin parameter \cite{pad}
\begin{equation}
\lambda = \frac{\omega}{\omega_{sup}} = \frac{{L} 
|E|^{1/2}}{G M^{5/2}}\, , 
\label{e2}
\end{equation}
that is, the ratio between the actual angular 
frequency $\omega$ of the system and the hypothetical angular 
frequency $\omega_{sup}$ needed to support the system against gravity 
purely by rotation. Here $|E| \simeq G M^{2} R^{-1}$ 
is the binding energy of the system, 
where $R$ is the radius and $M$ is the total mass of the 
protogalaxy. 
From simulations \cite{ban} it is obtained that the median value 
of $\lambda$ ($\lambda_{\text{med}} \sim 0.05$) for collapsed 
objects is insensitive 
to the shape of the initial power spectrum of density fluctuations 
or the magnitude of its initial overdensity. 
Spiral galaxies indicate an 
observed value of the spin parameter 
$\lambda_{0} \sim 0.5$ \cite{pad}. It is necessary thus to 
reconcile the angular momentum due to tidal torques 
$\lambda_{\text{med}} \sim 0.05$ 
with the observed value. It is accomplished 
considering the existence of a halo of dark matter so that 
the increase of the spin parameter is due to the increase of the 
binding energy $E$ in Eq.(\ref{e2}) because of the collapse 
\cite{pad}. In this process angular momentum remains constant, 
i.e., the angular momentum acquired up to the time protogalaxies 
became far apart (protogalactic decoupling time) is conserved.

We assume that the protogalaxy had a total mass 
$M \sim 10^{13} M_{\odot}$ corresponding to a large spiral galaxy
possessing a halo of dark matter 
$\sim 10$ times the mass of the luminous matter 
$M_{\text{L}} \sim 10^{12} M_{\odot}$. We also assume that the angular 
momentum of a protogalaxy increased, 
until the protogalaxies became sufficiently far apart and decoupled 
from the other protogalaxies, preserving their angular momentum, 
${\bf L}$, acquired from the tidal interaction with the other 
protogalaxies. 
We do not know when protogalaxies decoupled; we thus consider 
the decoupling redshifts $z_{d} = 100, 10, 5, 2, 0.5$, and $\sim 0$. 
We assume that up to the time of protogalaxy decoupling the mean 
density of the protogalaxy was roughly that of the Universe 
${\rho}(z) = (1 + z)^{3} \rho_{0}$, 
where $\rho_{0}$ is the present matter density of the Universe 
($\rho_{0} \sim 1.057 \times 10^{-29}\;\text{g}\,\text{cm}^{-3}$ 
with $H_{0} = 75\;\text{km}\,\text{s}^{-1}\,\text{Mpc}^{-1}$). 
The radius of the protogalaxy ${R}(z)$ is then 
${R}(z) = [(3/4 \pi)\, M\,  {\rho}(z)^{-1}]^{1/3}$. 
The angular momentum ${L}(z)$ 
acquired by a protogalaxy is 
\begin{equation}
{L}(z) \cong \frac{2}{5} \lambda_{\text{med}} \left[ G M^{3} 
{R}(z)\right]^{1/2}\, .
\label{e3}
\end{equation}

As noted above, we assume that 
${L}(z)$ in (\ref{e3}) and consequently ${\bf m}(z)$ 
in (\ref{e1}) are conserved after the decoupling redshift $z_{d}$. 
The magnetic field in the vicinity of the protogalaxy is obtained 
approximately through the relation 
${{\bf B}_{\text{NMC}}}(z) \cong {\bf m}(z) {R}(z)^{-3}$. 
Assuming that the magnetic field of the magnetic dipole is frozen 
into the plasma part of the galaxy which collapses to a present radius 
$R_{\text{L}} \simeq 10\;\text{kpc} \simeq 3.1 \times 10^{22}\;
\text{cm}$, 
we obtain a present magnetic field at the radius $R_{\text{L}}$, 
${\bf B}_{0}(R_{\text{L}}, z_{d}) \cong 
{{\bf B}_{\text{NMC}}}(z_{d}) [{R}(z_{d}) / R_{\text{L}}]^{3}$. 
Assuming a protogalaxy total mass $M \sim 10^{13} M_{\odot}$, 
we present in Table \ref{tab}, at the 
decoupling redshift $z_{d}$, the density of the protogalaxy, 
${\rho}(z_{d})$ ($\sim$ ambient density), the radius of the 
protogalaxy ${R}(z_{d})$, the magnetic field 
${B_{\text{NMC}}}(z_{d})$ taking $R \sim {R}(z_{d})$, 
the angular momentum ${L}(z_{d})$, the magnetic dipole 
moment ${m}(z_{d})$, and the present magnetic field 
${B_{0}}(R_{\text{L}}, z_{d})$. 
We have to take into account an additional amplification due to the 
differential rotation which ranges from 10 to 100. 
 
In Table \ref{tab} we use for $\beta$ in Eq.(\ref{e1}) $\beta = 
0.1$. We also take into account 
an additional amplification due to differential rotation 
of order $10$. Hence, we obtain for a decoupling redshift 
$z_{d} \alt 10$ the present magnetic field 
\begin{equation}
{{B}_{0}}(R_{\text{L}}, z_{d}) \sim 
10^{-6} \text{---} 10^{-5}\;\text{G}\, .
\label{e4}
\end{equation}

As noted above, we suggest a value for $\beta$ for galaxies 0.01 to 
0.1, which is consistent with the observational data of Sirag \cite{sir} 
and Woodward \cite{woo}. For $\beta$ equals to 0.01 with an 
amplification due to differential rotation of 100, we obtain the same 
values of ${B_{0}}(R_{\text{L}}, z_{d})$ as given 
in relation (\ref{e4}). We 
note that for a value of $\beta$ on the order of unity or greater than 
unity, the field predicted by NMC mechanism becomes inconsistent with 
the observations.

We assume that this poloidal field (\ref{e4}) is transformed into 
a toroidal field by the differential rotation of the spiral 
galaxy. However, it is possible that a large scale dynamo have 
influenced only the geometry more than the strength of 
magnetic fields \cite{kro}, so transforming poloidal fields 
into toroidal fields. Nonetheless, 
we note from (\ref{e4}) that no appreciable 
dynamo action is necessary to explain the presently observed 
magnetic field strength for decoupling redshifts $z_{d} \alt 10$.

For a galaxy of total mass less than $10^{13} M_{\odot}$ 
we would have a smaller magnetic field. 
We note from Table \ref{tab} that we have an estimate for the 
intergalactic magnetic field (the field between protogalaxies) 
for $z_{d} \alt 2$, 
${{B}_{\text{NMC}}}(z_{d}) \sim 10^{-14} \text{---} 
10^{-12}\;\text{G}$. 
This is consistent with our knowledge of the intergalactic 
magnetic field \cite{kro}.

The above discussion was for the origin of magnetic fields in 
spiral galaxies. We assume that the origin of magnetic fields in 
other types of galaxies is due to the merger of spiral galaxies or 
the diffusion of the magnetic field out of spiral galaxies.

R.O. would like to acknowledge the partial support of the Brazilian 
agency CNPq; U.F.W. the partial support of the Brazilian agency 
FAPESP and US Department of Energy under grant DE-F602-91ER40688, 
Task A.
U.F.W. would like to thank Dr. R. Brandenberger for the warm 
reception at Brown University.

\break

\begin{table}
\caption{Density ($\rho(z_{d})$), radius ($R(z_{d})$), magnetic field 
(${{B}_{NMC}}(z_{d})$), angular momentum (${L}(z_{d})$), 
magnetic moment (${m}(z)$) at the decoupling redshift ($z_{d}$), 
and the present magnetic field 
${{B}_{0}}(R_{\text{L}}, z_{d})$ at a radius 
$R_{\text{L}} \simeq 10\;\text{kpc} \simeq 3.1 \times 10^{22}\;
\text{cm}$ 
for a protogalaxy of total mass $M \sim 10^{13} M_{\odot}$.}
\label{tab}
\begin{tabular}{ccccccc}
$z_{d}$ & $\rho(z_{d})\;\;[\text{g}\,\text{cm}^{-3}]$ & 
$R(z_{d})\;\;[\text{cm}]$ & ${{B}_{NMC}}(z_{d})\;\;[\text{G}]$ & 
${L}(z_{d})\;\;[\text{g}\,\text{cm}^{2}\,\text{s}^{-1}]$ & 
${m}(z)\;\;[\text{erg}\,\text{G}^{-1}]$ & 
${{B}_{0}}(R_{\text{L}}, z_{d})\;\;[\text{G}]$ \\  \hline 
100 & $1.1 \times 10^{-23}$ & $7.6 \times 10^{22}$ & $4.0 \times 
10^{-9}$ & $4.0 \times 10^{75}$ & $1.7 \times 
10^{61}$ & $5.8 \times 10^{-7}$   \\ 
10 & $1.4 \times 10^{-26}$ & $7.0 \times 10^{23}$ & $1.5 \times 
10^{-11}$ & $1.2 \times 10^{76}$ & $5.2 \times 
10^{61}$ & $1.8 \times 10^{-6}$   \\ 
5 & $2.3 \times 10^{-27}$ & $1.3 \times 10^{24}$ & $3.4 \times 
10^{-12}$ & $1.6 \times 10^{76}$ & $7.0 \times 
10^{61}$ & $2.4 \times 10^{-6}$     \\ 
2 & $2.8 \times 10^{-28}$ & $2.5 \times 10^{24}$ & $6.0 \times 
10^{-13}$ & $2.3 \times 10^{76}$ & $10.0 \times 
10^{61}$ & $3.4 \times 10^{-6}$    \\ 
0.5 & $3.6 \times 10^{-29}$ & $5.1 \times 10^{24}$ & $1.0 \times 
10^{-13}$ & $3.3 \times 10^{76}$ & $1.4 \times 
10^{62}$ & $4.8 \times 10^{-6}$   \\ 
$ \sim 0$ & $1.0 \times 10^{-29}$ & $7.6 \times 10^{24}$ & $3.8 \times 
10^{-14}$ & $4.0 \times 10^{76}$ & $1.7 \times 
10^{62}$ & $5.9 \times 10^{-6}$ 
\end{tabular}
\end{table}

\end{document}